\documentclass[apl,aps,twocolumn,epsfig]{revtex4}
\usepackage{graphicx}
\usepackage{dcolumn}
\usepackage{bm}
\usepackage{threeparttable}
\usepackage{gensymb}
\makeatletter

\newcommand{\Rmnum}[1]{\expandafter\@slowromancap\romannumeral #1@}
\makeatother
\begin{document}
\title {Influence of etching processes on electronic transport in mesoscopic InAs/GaSb quantum well devices}
\author{Atindra Nath Pal, Susanne M\"uller, Thomas Ihn, Klaus Ensslin, Thomas Tschirky, Christophe Charpentier, and Werner Wegscheider} \vspace{1.5cm}
\address{Solid State Physics Laboratory, ETH Z\"urich - 8093 Z\"urich, Switzerland}

\begin{abstract}
We report the electronic characterization of mesoscopic Hall bar devices fabricated from coupled InAs/GaSb quantum wells sandwiched between AlSb barriers, an emerging candidate for two-dimensional topological insulators. The electronic width of the etched structures was determined from the low field magneto-resistance peak, a characteristic signature of partially diffusive boundary scattering in the ballistic limit. In case of dry-etching the electronic width was found to decrease with electron density. In contrast, for wet etched devices it stayed constant with density. Moreover, the boundary scattering was found to be more specular for wet-etched devices, which may be relevant for studying topological edge states.
\end{abstract}


\maketitle
\section{Introduction}
Recently, InAs/GaSb composite quantum wells have gained a lot of interest due to the prediction of the quantum spin Hall (QSH) state in the inverted regime~\cite{QSHE_Liu_InAs_GaSb} of the bandstructure. In this state, transport is expected to be governed by counter propagating (helical) edge channels of opposite spins together with an insulating bulk and the system is referred to as a two-dimensional topological insulator (2D TI). The QSH state was first predicted~\cite{Bernevig_Zhang_science} and observed~\cite{Science_Molenkamp} in HgTe/CdTe quantum wells. Electron-hole hybridization and the indication of a topological insulator phase were reported in InAs/GaSb quantum wells in recent transport experiments~\cite{Knez_PRB_2010,Knez_PRL_2011,Muraki_PRB2013,RRDu_arxiv2013,PRL_Fabrizio,Knez_PRL2014}. One important requirement to observe the 2D TI phase in electronic transport is to fabricate devices smaller than the inelastic scattering length (l$_\mathrm{in}$). In reality, due to limited material quality often bulk transport is relevant and masks edge channel transport~\cite{Knez_PRL_2011}. Moreover, edge scattering originating from rough edges as a result of fabrication processes, can also be detrimental to the observation of helical edge states~\cite{scanning_gate_Gordan,Glazman_PRL13,edge_scattering_arxiv14}. Apart from being an emerging candidate for 2D TI, this system offers the potential to observe several physical phenomena such as exciton condensation~\cite{exciton_Pikulin_PRL14}, Majorana Fermions~\cite{Beenakker_Majorana}, or edge mode superconductivity~\cite{Vlad_arxive_2014}. In comparison to the HgTe/CdTe system, InAs/GaSb coupled quantum wells offer electric field tunability of the topological phase~\cite{Naveh_PRL}.

\begin{figure}[!t]
\begin{center}
\includegraphics [width=1\linewidth]{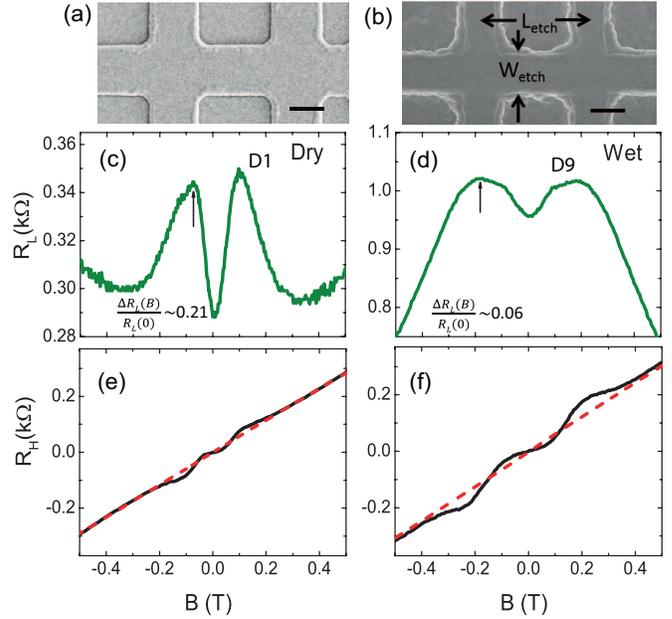}
\end{center}
\caption{ SEM images of typical (a) dry and (b) wet-etched devices. Arrows in (b) indicate the length between the voltage probes ($L_\mathrm{etch}$) and the width ($W_\mathrm{etch}$) of the Hall bar after etching. Scale bar is $2~\mu$m. (c)-(d) Longitudinal four-terminal resistance ($R_\mathrm{L}$) and (e)-(f) (black solid lines) the Hall resistance ($R_H$), as a function of magnetic field (B) for devices D1 (dry-etched) and D9 (wet-etched), respectively, at a density $n\approx 9.5\times 10^{11}$~cm$^{-2}$. The red dotted lines in (e) and (f) are $R_H-B$ curves at similar electronic density for wider devices, D3 (W=25~$\mu$m, dry-etched) and D11 (W=22.2~$\mu$m, wet-etched), respectively, indicating the absence of a quenched Hall effect near $B=0$. The arrows in (c) and (d) indicate $B_{max}$ (see text). } \label{fig1}
\end{figure}

\begin{table*}
Table I: Details of the Hall bar devices: $W_\mathrm{etch}$ and $L_\mathrm{etch}$ are respectively the measured width and length between the voltage probes after etching. $W_\mathrm{electronic}$ is the electronic width extracted from the measurements.
\vspace{0.4 cm}
\label{Tab1}
\begin{threeparttable}[t]
\begin{tabular}{c c c c c c c c c}\hline
 Device & Wafer & Etching & Etching depth & $W_\mathrm{etch}$ & $L_\mathrm{etch}$ &$W_\mathrm{electronic}$\tnote{1}& $\Delta R_L(B)/R_L(0)$\tnote{1} & Remark\\
  &  & type & [nm] & [$\mu$m] & [$\mu$m] &\\
  \hline
 &\\
 D1 & W1 & Dry & 300 & $2.0 \pm 0.10$ & $10.0$ & 0.90& 0.21 &  \\
 D2 & W1 & Dry & 300 & $4.1 \pm 0.10$ & $10.0$ & 1.80 & 0.06&  \\
 D3 & W1 & Dry & 300 & $25.0 \pm 0.10$ & $50.0$ & - & - & No detectable MR peak \\
 D4 & W3 & Dry & 300 & $2.1 \pm 0.10$ & $12.0$ & 0.85 & 0.17&  \\
 D5 & W3 & Dry & 300 & $4.1 \pm 0.10$ & $18.0$ & - &- & No detectable MR peak\\
 D6 & W2 & Wet & 200 & $1.9 \pm 0.17$ & $11.5$ & 0.75 & 0.02& \\
 D7 & W2 & Wet & 350 & $2.6 \pm 0.20$ & $10.0$ & 1.10 & 0.00&  \\
 D8 & W2 & Wet & 350 & $4.4 \pm 0.25$ & $10.0$ & - & - & No detectable MR peak \\
 D9 & W3 & Wet & 1100 & $1.7 \pm 0.42$ & $10.0$ & 0.45 & 0.06& \\
 D10 & W3 & Wet & 240 & $0.9 \pm 0.12$ & $11.0$ & 0.46 & 0.01 &  \\
 D11 & W3 & Wet & 1600 & $22.2 \pm 0.20$ & $50.0$ & - & - & No detectable MR peak \\

 &\\
 \hline
\end{tabular}
\begin{tablenotes}
    \item[1] for $n\approx 9.5\times 10^{11}$ cm$^{-2}$
\end{tablenotes}
\end{threeparttable}

\end{table*}

When the device size becomes smaller than the elastic mean free path ($l_\mathrm{e}$), the transport properties are modified with respect to bulk samples~\cite{Theory_size_effect66,Theory_1d_channel,Beenakker_Houten_prl89}.  As a consequence mesoscopic devices exhibit enhanced resistivity due to the lateral boundary scattering, studied extensively in narrow channels fabricated in GaAs two-dimensional electron gases (2DEGs)~\cite{Boundary_scat_Thornton_89,Quenching_FORD_PRL89}. The electronic confinement in narrow 2DEGs depends a lot on the material system and particularly on the fabrication processes. For example, quasi-1D channels fabricated by reactive ion etching (RIE) in InAs 2DEGs~\cite{APL_InAs_RIE} show a degradation of the mobility due to surface damage caused by the energetic ions along with lateral side wall depletion. Electronic transport in sub-micron devices of InAs~\cite{InAS_PRB_2002} and InSb~\cite{InSb_PRB2011} based 2D systems were studied extensively. However, mesoscopic transport in InAs/GaSb systems is relatively unexplored. Compared to the pure InAs well, the presence of a GaSb layer may change the depletion width and the confinement potential. Hence, it is necessary to explore the mesoscopic properties of InAs/GaSb devices for reaching an optimized fabrication recipe.

Here, we present a detailed characterization of a series of mesoscopic Hall bar devices made by dry and wet etching processes (see Table~I for details). The effective electronic width ($W_\mathrm{electronic}$) was determined by measuring the low field magneto-resistance peak appearing due to partially diffusive boundary scattering~\cite{Theory_size_effect66}. We find that $W_\mathrm{electronic}$ decreases with decreasing density for dry-etched devices, whereas it remains constant for wet-etched devices. The edge roughness in wet-etched devices was found to be smaller compared to dry-etched devices, which could be relevant for observing edge transport in this material.

\section{Fabrication}
The devices were fabricated from wafers grown by molecular-beam epitaxy (MBE) with a similar growth structure as in Ref.~\onlinecite{PRL_Fabrizio}. We present measurements on eleven Hall bar devices from three different wafers (W1, W2 and W3) with exactly the same nominal growth structure, but with different targeted lateral width. For the fabrication of the Hall bar devices, Ohmic contacts Ge(18~nm)/Au($50$~nm)/Ni($40$~nm)/Au($100$~nm) and mesa were defined by optical lithography. For dry-etching, an Inductively Coupled Plasma (ICP) of Ar was used with a flow rate of $20$~sccm. For wet etching we have used a phosphoric acid and the citric acid based III-V etchant, H$_3$PO$_4$:H$_2$O$_2$:C$_6$H$_8$O$_7$:H$_2$O ($3:5:55:220$) with a nominal etching rate of $10-15$~nm/minute~\cite{Knez_thesis}. After etching, the devices are passivated with $200$~nm of Silicon Nitride, deposited by Plasma Enhanced Chemical Vapor Deposition (PECVD) at $\sim 300$~\degree C, which serves also as the dielectric for the top gate. Finally, the top gate is defined by one more step of optical lithography followed by the deposition of Ti/Au and lift-off. Figure~1a and 1b show the SEM images of typical dry-etched and wet-etched devices, respectively. Arrows in figure~1b indicate the measured length between the voltage probes ($L_\mathrm{etch}$) and width ($W_\mathrm{etch}$) after etching for the wet-etched device.

\section{Results and Discussion}
We found all of our wafers to be electron doped at zero gate voltage with a density $n\approx 7\times 10^{11}$~cm$^{-2}$ irrespective of the fabrication processes. All the experiments were done by ac lock-in technique with a bias current of $10-50$~nA and carrier frequency of 31~Hz at $1.3$~K. Both electron and hole transport can be observed by tuning the gate voltage from positive to negative. Transport measurements on large Hall bars ($25~\mu$m$\times50~\mu$m) fabricated by both dry and wet etching possess similar carrier-mobilities and gate tunability as in Ref.~\onlinecite{PRL_Fabrizio}.

\begin{figure*}[!htb]
\begin{center}
\includegraphics [width=1\linewidth]{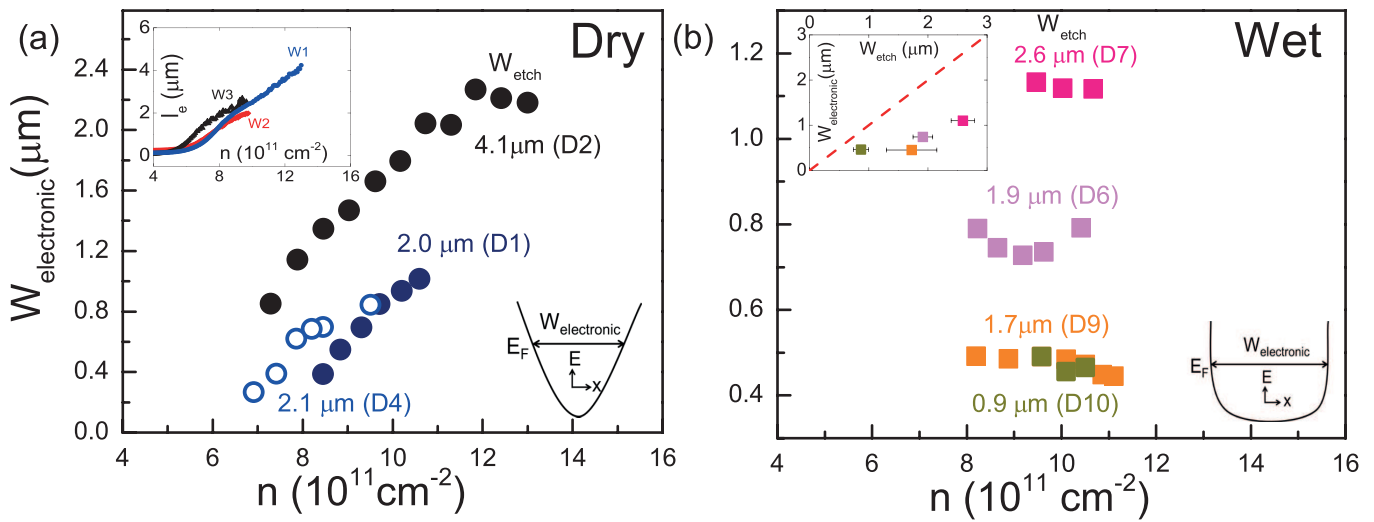}
\end{center}
\caption{ The electronic width (W$_{electronic}$) is plotted as a function of density ($n$) for both (a) dry and (b) wet etched devices. The etched width ($W_\mathrm{etch}$) and the corresponding device name are indicated for each devices. (Top left inset in a) Mean free path (l$_\mathrm{e}$) is plotted as a function of density for three different wafers (W1, W2, W3). (Top left inset in b) Average $W_\mathrm{electronic}$ is plotted against the mean etched width ($W_\mathrm{etch}$) for wet-etched devices. The error bar indicates the uncertainty in measuring $W_\mathrm{etch}$. The dashed line corresponds to $W_\mathrm{electronic}=W_\mathrm{etch}$. (Bottom insets) The electronic confinement potential for both etching processes is presented  schematically. } \label{fig2}
\end{figure*}

Boundary scattering was characterized by measuring the longitudinal resistance ($R_\mathrm{L}$) and its magnetic field dependence~\cite{Boundary_scat_Thornton_89}. Figure~1c and 1d show typical low-field magneto-resistance (MR) data for dry and wet etched devices (devices~D1 and D9 in Table~I) at an electronic density, $n\approx 9.5\times 10^{11}$ cm$^{-2}$. We find distinct MR peaks which are used to determine the electronic width of the devices. When the device width becomes smaller than the elastic mean free path ($l_{e}$), charge carriers face partially diffusive boundary scattering in long channels ($L>l_\mathrm{e}$). This results in a positive zero-field MR reaching a maximum at a field $B_{max}$ (indicated by arrows in figure~1c and 1d) when the ratio of the wire width ($W_\mathrm{electronic}$) and the cyclotron radius ($R_c$) follows the relation~\cite{Theory_size_effect66} $W_\mathrm{electronic}/R_\mathrm{c}=W_\mathrm{electronic}eB_\mathrm{max}/\hbar k_\mathrm{F}=0.55$, where $k_\mathrm{F}=(2\pi n)^{1/2}$, is the Fermi wave vector. In the top left inset of figure~2a, we compare $l_e$ as a function of electronic density ($n$) for the three wafers (W1, W2 and W3), calculated from wider devices (D3, D8 and D11) where MR peaks were not observed. $l_e$ was calculated using the relation, $l_e=h/\rho e^2 k_F$~\cite{Thomas_book}, $\rho$ being the measured resistivity.

Figure~1e and 1f (black solid lines) show the transverse MR ($R_\mathrm{H}$) at $n\approx 9.5\times 10^{11}$ cm$^{-2}$ for devices~D1 and D9, respectively, exhibiting quenching of the Hall resistance near zero field, a typical classical phenomenon of electron transport in ballistic crosses~\cite{Quenching_FORD_PRL89,Beenakker_Houten_prl89}. For wider devices the Hall effect becomes linear with magnetic field, indicated by the red dotted lines in figure~1e and 1f, obtained from devices D3 (W=25~$\mu$m, dry-etched) and D11 (W=22.2~$\mu$m, wet-etched), respectively. Boundary scattering was observed mainly for high mobility electrons as $l_\mathrm{e}$ exceeds the device dimension only for densities larger than $\sim 7\times 10^{11}$ cm$^{-2}$. No ballistic effects were observed for holes in this temperature range due to the low mobility and high effective mass. The roughness of the boundary can qualitatively be evaluated by calculating the quantity $\Delta R_\mathrm{L}(B)/R_\mathrm{L}(0)$~\cite{Boundary_scat_Thornton_89}, where, $\Delta R_\mathrm{L}(B)=R_\mathrm{L}(B_\mathrm{max})-R_\mathrm{L}(0)$, which is zero for completely specular boundary scattering. From figure~1c and 1d, we see that $\Delta R_\mathrm{L}(B)/R_\mathrm{L}(0)$ is $\sim 0.21$ for device~D1 and $\sim 0.06$ for device~D9, confirming the smoother edge for the wet etched device (see Table~I for other devices).

Next we concentrate on the density ($n$) dependence of the electronic width ($W_\mathrm{electronic}$). Figure~2a shows the density dependence of three dry-etched devices (devices D1, D2, D4) with mean $W_\mathrm{etch}$ $2.0~\mu$m, $4.1~\mu$m and $2.1~\mu$m, respectively. For all the three devices, $W_\mathrm{electronic}$ was found to decrease with decreasing electron density. For device~D2 ($W_\mathrm{etch}\approx 4.1~\mu$m), $W_\mathrm{electronic}$ saturates at around $2.2~\mu$m above a density $n\approx 1.1\times 10^{12}$ cm$^{-2}$ and decreases to $\approx 0.8~\mu$m at  $n\approx 7.3\times 10^{11}$ cm$^{-2}$, the lowest density where the MR peak was observed. Below this density the holes start populating the GaSb layer and transport is governed by both electrons and holes, where we see the signature of two band transport in both longitudinal and Hall resistance~\cite{Fabrizio_thesis}.  For devices~D1 and D4 ($W_\mathrm{etch}\approx 2.1~\mu$m), the saturation was not observed due to the limited gate range, however, a similar decrease in $W_\mathrm{electronic}$ was observed.  Figure~2b shows the density dependence of $W_\mathrm{electronic}$ for all the wet-etched devices with a mean $W_\mathrm{etch}$ ranging between $0.9-2.6~\mu$m. For devices with $W_\mathrm{etch}\approx 2.6~\mu$m (device~D7) and $\approx 1.9~\mu$m (device~D6), $W_\mathrm{electronic}$ was found to be $1.1~\mu$m and $0.75~\mu$m, respectively, with little variation with density. For devices D9 ($W_\mathrm{etch}\approx 1.7~\mu$m) and D10 ($W_\mathrm{etch}\approx 0.9~\mu$m), $W_\mathrm{electronic}$ shows a similar  value $\sim 0.5~\mu$m. Due to the larger etching depth for device D9 ($\approx 1.1~\mu$m) the side wall etching is expected to be more compared to device~D10 (etching depth $\approx 240$~nm) and may provide coincidentally similar $W_\mathrm{electronic}$.

The density dependence of $W_{electronic}$ (figure~2) suggests a steeper confinement potential for wet-etched devices compared to the dry-etched ones (see insets of figure~2a and 2b), however, the exact shape of the confinement potential is difficult to model.

The depletion width is often used in literature to judge the quality of etching processes, defined as $W_{depletion}=1/2(W_\mathrm{etch}-W_\mathrm{electronic})$. As $W_\mathrm{electronic}$ varies with density for dry-etched devices, the depletion width is also not constant unlike for wet-etched devices, where $W_\mathrm{electronic}$ remains constant with density. To get a reasonable comparison for the depletion width values, we have plotted average $W_\mathrm{electronic}$ as a function of $W_\mathrm{etch}$ only for the wet-etched devices in the top inset of figure~2b. The red dashed line indicates the line corresponding to $W_\mathrm{electronic} = W_\mathrm{etch}$. In wet etching we often find double step and corrugation at the edges due to the different etching rates for layers with different materials, in contrary to the dry etching where the etching is uniform (see figure~1a and 1b). This may lead to a larger uncertainty in determining the actual $W_\mathrm{etch}$ for the wet-etched devices. We find $W_\mathrm{depletion}$ to be in the range $\sim 0.2-0.8~\mu$m, which is rather large compared to the pure InAs quantum well devices~\cite{InAS_PRB_2002}. Moreover, the lateral depletion was found to increase for wider devices, which is counterintuitive. We also looked at the $W_\mathrm{depletion}$ as a function of etching depth. However, no systematic dependence of $W_\mathrm{depletion}$ with etching depth was observed in these four devices.

Qualitatively, in these devices we found that the gate capacitance per unit area ($e.dn/dV_{TG}$) decreases for narrower devices. For a narrow channel with an ideal edge, one expects to have enhanced gate tunability due to the fringing field lines near the edge. We speculate that due to the etching process there could be depositions of non-volatile components, ionic charges \emph{etc.},~\cite{Wet_etching_superlattice}. These etchant residues may create charge traps near the edges which may be populated by the application of gate voltage. These trapped charges may screen the gate voltage, as well as modify the confinement potential in the lateral direction.

We plan to look into the influence of etching processes on the local and non-local electronic transport properties. Further work is necessary to reduce the bulk conduction either by improving the material quality or impurity doping~\cite{RRDu_arxiv2013} to observe the quantum spin Hall effect in this system.

\section{Conclusion}
In conclusion, we have been able to fabricate and characterize mesoscopic devices of InAs/GaSb quantum well, made by two different etching processes. The electronic width was obtained from the low field magneto-resistance peak appearing due to boundary scattering. For dry-etched devices the electronic width decreases with lowering electron density, whereas, it remains constant for wet-etched devices. Our observations suggest that wet chemical etching is less invasive compared to the dry-etching and could be preferable to produce mesoscopic devices for studying topological insulators.

\section{Acknowledgements}
The authors wish to thank the Swiss National Science Foundation for financial support via NCCR QSIT (Quantum Science and Technology).


\begin{thebibliography}{28}
\expandafter\ifx\csname natexlab\endcsname\relax\def\natexlab#1{#1}\fi
\expandafter\ifx\csname bibnamefont\endcsname\relax
  \def\bibnamefont#1{#1}\fi
\expandafter\ifx\csname bibfnamefont\endcsname\relax
  \def\bibfnamefont#1{#1}\fi
\expandafter\ifx\csname citenamefont\endcsname\relax
  \def\citenamefont#1{#1}\fi
\expandafter\ifx\csname url\endcsname\relax
  \def\url#1{\texttt{#1}}\fi
\expandafter\ifx\csname urlprefix\endcsname\relax\def\urlprefix{URL }\fi
\providecommand{\bibinfo}[2]{#2}
\providecommand{\eprint}[2][]{\url{#2}}

\bibitem[{\citenamefont{Liu et~al.}(2008)\citenamefont{Liu, Hughes, Qi, Wang,
  and Zhang}}]{QSHE_Liu_InAs_GaSb}
\bibinfo{author}{\bibfnamefont{C.}~\bibnamefont{Liu}},
  \bibinfo{author}{\bibfnamefont{T.~L.} \bibnamefont{Hughes}},
  \bibinfo{author}{\bibfnamefont{X.-L.} \bibnamefont{Qi}},
  \bibinfo{author}{\bibfnamefont{K.}~\bibnamefont{Wang}}, \bibnamefont{and}
  \bibinfo{author}{\bibfnamefont{S.-C.} \bibnamefont{Zhang}},
  \bibinfo{journal}{Phys. Rev. Lett.} \textbf{\bibinfo{volume}{100}},
  \bibinfo{pages}{236601} (\bibinfo{year}{2008}).

\bibitem[{\citenamefont{Bernevig et~al.}(2006)\citenamefont{Bernevig, Hughes,
  and Zhang}}]{Bernevig_Zhang_science}
\bibinfo{author}{\bibfnamefont{B.~A.} \bibnamefont{Bernevig}},
  \bibinfo{author}{\bibfnamefont{T.~L.} \bibnamefont{Hughes}},
  \bibnamefont{and} \bibinfo{author}{\bibfnamefont{S.-C.} \bibnamefont{Zhang}},
  \bibinfo{journal}{Science} \textbf{\bibinfo{volume}{314}},
  \bibinfo{pages}{1757} (\bibinfo{year}{2006}).

\bibitem[{\citenamefont{K\"onig et~al.}(2007)\citenamefont{K\"onig, Wiedmann,
  Brüne, Roth, Buhmann, Molenkamp, Qi, and Zhang}}]{Science_Molenkamp}
\bibinfo{author}{\bibfnamefont{M.}~\bibnamefont{K\"onig}},
  \bibinfo{author}{\bibfnamefont{S.}~\bibnamefont{Wiedmann}},
  \bibinfo{author}{\bibfnamefont{C.}~\bibnamefont{Brüne}},
  \bibinfo{author}{\bibfnamefont{A.}~\bibnamefont{Roth}},
  \bibinfo{author}{\bibfnamefont{H.}~\bibnamefont{Buhmann}},
  \bibinfo{author}{\bibfnamefont{L.~W.} \bibnamefont{Molenkamp}},
  \bibinfo{author}{\bibfnamefont{X.-L.} \bibnamefont{Qi}}, \bibnamefont{and}
  \bibinfo{author}{\bibfnamefont{S.-C.} \bibnamefont{Zhang}},
  \bibinfo{journal}{Science} \textbf{\bibinfo{volume}{318}},
  \bibinfo{pages}{766} (\bibinfo{year}{2007}).

\bibitem[{\citenamefont{Knez et~al.}(2010)\citenamefont{Knez, Du, and
  Sullivan}}]{Knez_PRB_2010}
\bibinfo{author}{\bibfnamefont{I.}~\bibnamefont{Knez}},
  \bibinfo{author}{\bibfnamefont{R.~R.} \bibnamefont{Du}}, \bibnamefont{and}
  \bibinfo{author}{\bibfnamefont{G.}~\bibnamefont{Sullivan}},
  \bibinfo{journal}{Phys. Rev. B} \textbf{\bibinfo{volume}{81}},
  \bibinfo{pages}{201301} (\bibinfo{year}{2010}).

\bibitem[{\citenamefont{Knez et~al.}(2011{\natexlab{a}})\citenamefont{Knez, Du,
  and Sullivan}}]{Knez_PRL_2011}
\bibinfo{author}{\bibfnamefont{I.}~\bibnamefont{Knez}},
  \bibinfo{author}{\bibfnamefont{R.-R.} \bibnamefont{Du}}, \bibnamefont{and}
  \bibinfo{author}{\bibfnamefont{G.}~\bibnamefont{Sullivan}},
  \bibinfo{journal}{Phys. Rev. Lett.} \textbf{\bibinfo{volume}{107}},
  \bibinfo{pages}{136603} (\bibinfo{year}{2011}{\natexlab{a}}).

\bibitem[{\citenamefont{Suzuki et~al.}(2013)\citenamefont{Suzuki, Harada,
  Onomitsu, and Muraki}}]{Muraki_PRB2013}
\bibinfo{author}{\bibfnamefont{K.}~\bibnamefont{Suzuki}},
  \bibinfo{author}{\bibfnamefont{Y.}~\bibnamefont{Harada}},
  \bibinfo{author}{\bibfnamefont{K.}~\bibnamefont{Onomitsu}}, \bibnamefont{and}
  \bibinfo{author}{\bibfnamefont{K.}~\bibnamefont{Muraki}},
  \bibinfo{journal}{Phys. Rev. B} \textbf{\bibinfo{volume}{87}},
  \bibinfo{pages}{235311} (\bibinfo{year}{2013}).

\bibitem[{\citenamefont{{Du} et~al.}(2013)\citenamefont{{Du}, {Knez},
  {Sullivan}, and {Du}}}]{RRDu_arxiv2013}
\bibinfo{author}{\bibfnamefont{L.}~\bibnamefont{{Du}}},
  \bibinfo{author}{\bibfnamefont{I.}~\bibnamefont{{Knez}}},
  \bibinfo{author}{\bibfnamefont{G.}~\bibnamefont{{Sullivan}}},
  \bibnamefont{and} \bibinfo{author}{\bibfnamefont{R.-R.} \bibnamefont{{Du}}},
  \bibinfo{journal}{ArXiv e-prints}  (\bibinfo{year}{2013}),
  \eprint{1306.1925}.

\bibitem[{\citenamefont{Nichele et~al.}(2014)\citenamefont{Nichele, Pal,
  Pietsch, Ihn, Ensslin, Charpentier, and Wegscheider}}]{PRL_Fabrizio}
\bibinfo{author}{\bibfnamefont{F.}~\bibnamefont{Nichele}},
  \bibinfo{author}{\bibfnamefont{A.~N.} \bibnamefont{Pal}},
  \bibinfo{author}{\bibfnamefont{P.}~\bibnamefont{Pietsch}},
  \bibinfo{author}{\bibfnamefont{T.}~\bibnamefont{Ihn}},
  \bibinfo{author}{\bibfnamefont{K.}~\bibnamefont{Ensslin}},
  \bibinfo{author}{\bibfnamefont{C.}~\bibnamefont{Charpentier}},
  \bibnamefont{and}
  \bibinfo{author}{\bibfnamefont{W.}~\bibnamefont{Wegscheider}},
  \bibinfo{journal}{Phys. Rev. Lett.} \textbf{\bibinfo{volume}{112}},
  \bibinfo{pages}{036802} (\bibinfo{year}{2014}).

\bibitem[{\citenamefont{Knez et~al.}(2011{\natexlab{b}})\citenamefont{Knez, Du,
  and Sullivan}}]{Knez_PRL2014}
\bibinfo{author}{\bibfnamefont{I.}~\bibnamefont{Knez}},
  \bibinfo{author}{\bibfnamefont{R.-R.} \bibnamefont{Du}}, \bibnamefont{and}
  \bibinfo{author}{\bibfnamefont{G.}~\bibnamefont{Sullivan}},
  \bibinfo{journal}{Phys. Rev. Lett.} \textbf{\bibinfo{volume}{107}},
  \bibinfo{pages}{136603} (\bibinfo{year}{2011}{\natexlab{b}}).

\bibitem[{\citenamefont{K\"onig et~al.}(2013)\citenamefont{K\"onig, Baenninger,
  Garcia, Harjee, Pruitt, Ames, Leubner, Br\"une, Buhmann, Molenkamp
  et~al.}}]{scanning_gate_Gordan}
\bibinfo{author}{\bibfnamefont{M.}~\bibnamefont{K\"onig}},
  \bibinfo{author}{\bibfnamefont{M.}~\bibnamefont{Baenninger}},
  \bibinfo{author}{\bibfnamefont{A.~G.~F.} \bibnamefont{Garcia}},
  \bibinfo{author}{\bibfnamefont{N.}~\bibnamefont{Harjee}},
  \bibinfo{author}{\bibfnamefont{B.~L.} \bibnamefont{Pruitt}},
  \bibinfo{author}{\bibfnamefont{C.}~\bibnamefont{Ames}},
  \bibinfo{author}{\bibfnamefont{P.}~\bibnamefont{Leubner}},
  \bibinfo{author}{\bibfnamefont{C.}~\bibnamefont{Br\"une}},
  \bibinfo{author}{\bibfnamefont{H.}~\bibnamefont{Buhmann}},
  \bibinfo{author}{\bibfnamefont{L.~W.} \bibnamefont{Molenkamp}},
  \bibnamefont{et~al.}, \bibinfo{journal}{Phys. Rev. X}
  \textbf{\bibinfo{volume}{3}}, \bibinfo{pages}{021003} (\bibinfo{year}{2013}).

\bibitem[{\citenamefont{V\"ayrynen et~al.}(2013)\citenamefont{V\"ayrynen,
  Goldstein, and Glazman}}]{Glazman_PRL13}
\bibinfo{author}{\bibfnamefont{J.~I.} \bibnamefont{V\"ayrynen}},
  \bibinfo{author}{\bibfnamefont{M.}~\bibnamefont{Goldstein}},
  \bibnamefont{and} \bibinfo{author}{\bibfnamefont{L.~I.}
  \bibnamefont{Glazman}}, \bibinfo{journal}{Phys. Rev. Lett.}
  \textbf{\bibinfo{volume}{110}}, \bibinfo{pages}{216402}
  (\bibinfo{year}{2013}).

\bibitem[{\citenamefont{{Entin} and
  {Magarill}}(2014)}]{edge_scattering_arxiv14}
\bibinfo{author}{\bibfnamefont{M.~V.} \bibnamefont{{Entin}}} \bibnamefont{and}
  \bibinfo{author}{\bibfnamefont{L.~I.} \bibnamefont{{Magarill}}},
  \bibinfo{journal}{ArXiv e-prints}  (\bibinfo{year}{2014}),
  \eprint{1407.3946}.

\bibitem[{\citenamefont{Pikulin and Hyart}(2014)}]{exciton_Pikulin_PRL14}
\bibinfo{author}{\bibfnamefont{D.~I.} \bibnamefont{Pikulin}} \bibnamefont{and}
  \bibinfo{author}{\bibfnamefont{T.}~\bibnamefont{Hyart}},
  \bibinfo{journal}{Phys. Rev. Lett.} \textbf{\bibinfo{volume}{112}},
  \bibinfo{pages}{176403} (\bibinfo{year}{2014}).

\bibitem[{\citenamefont{Mi et~al.}(2013)\citenamefont{Mi, Pikulin, Wimmer, and
  Beenakker}}]{Beenakker_Majorana}
\bibinfo{author}{\bibfnamefont{S.}~\bibnamefont{Mi}},
  \bibinfo{author}{\bibfnamefont{D.~I.} \bibnamefont{Pikulin}},
  \bibinfo{author}{\bibfnamefont{M.}~\bibnamefont{Wimmer}}, \bibnamefont{and}
  \bibinfo{author}{\bibfnamefont{C.~W.~J.} \bibnamefont{Beenakker}},
  \bibinfo{journal}{Phys. Rev. B} \textbf{\bibinfo{volume}{87}},
  \bibinfo{pages}{241405} (\bibinfo{year}{2013}).

\bibitem[{\citenamefont{{Pribiag} et~al.}(2014)\citenamefont{{Pribiag},
  {Beukman}, {Qu}, {Cassidy}, {Charpentier}, {Wegscheider}, and
  {Kouwenhoven}}}]{Vlad_arxive_2014}
\bibinfo{author}{\bibfnamefont{V.~S.} \bibnamefont{{Pribiag}}},
  \bibinfo{author}{\bibfnamefont{A.~J.~A.} \bibnamefont{{Beukman}}},
  \bibinfo{author}{\bibfnamefont{F.}~\bibnamefont{{Qu}}},
  \bibinfo{author}{\bibfnamefont{M.~C.} \bibnamefont{{Cassidy}}},
  \bibinfo{author}{\bibfnamefont{C.}~\bibnamefont{{Charpentier}}},
  \bibinfo{author}{\bibfnamefont{W.}~\bibnamefont{{Wegscheider}}},
  \bibnamefont{and} \bibinfo{author}{\bibfnamefont{L.~P.}
  \bibnamefont{{Kouwenhoven}}}, \bibinfo{journal}{ArXiv e-prints}
  (\bibinfo{year}{2014}), \eprint{1408.1701}.

\bibitem[{\citenamefont{Naveh and Laikhtman}(1996)}]{Naveh_PRL}
\bibinfo{author}{\bibfnamefont{Y.}~\bibnamefont{Naveh}} \bibnamefont{and}
  \bibinfo{author}{\bibfnamefont{B.}~\bibnamefont{Laikhtman}},
  \bibinfo{journal}{Phys. Rev. Lett.} \textbf{\bibinfo{volume}{77}},
  \bibinfo{pages}{900} (\bibinfo{year}{1996}).

\bibitem[{\citenamefont{{Ditlefsen} and {Lothe}}(1966)}]{Theory_size_effect66}
\bibinfo{author}{\bibfnamefont{E.}~\bibnamefont{{Ditlefsen}}} \bibnamefont{and}
  \bibinfo{author}{\bibfnamefont{J.}~\bibnamefont{{Lothe}}},
  \bibinfo{journal}{Philosophical Magazine} \textbf{\bibinfo{volume}{14}},
  \bibinfo{pages}{759} (\bibinfo{year}{1966}).

\bibitem[{\citenamefont{Berggren et~al.}(1988)\citenamefont{Berggren, Roos, and
  van Houten}}]{Theory_1d_channel}
\bibinfo{author}{\bibfnamefont{K.-F.} \bibnamefont{Berggren}},
  \bibinfo{author}{\bibfnamefont{G.}~\bibnamefont{Roos}}, \bibnamefont{and}
  \bibinfo{author}{\bibfnamefont{H.}~\bibnamefont{van Houten}},
  \bibinfo{journal}{Phys. Rev. B} \textbf{\bibinfo{volume}{37}},
  \bibinfo{pages}{10118} (\bibinfo{year}{1988}).

\bibitem[{\citenamefont{Beenakker and van
  Houten}(1989)}]{Beenakker_Houten_prl89}
\bibinfo{author}{\bibfnamefont{C.~W.~J.} \bibnamefont{Beenakker}}
  \bibnamefont{and} \bibinfo{author}{\bibfnamefont{H.}~\bibnamefont{van
  Houten}}, \bibinfo{journal}{Phys. Rev. Lett.} \textbf{\bibinfo{volume}{63}},
  \bibinfo{pages}{1857} (\bibinfo{year}{1989}).

\bibitem[{\citenamefont{Thornton et~al.}(1989)\citenamefont{Thornton, Roukes,
  Scherer, and Van~de Gaag}}]{Boundary_scat_Thornton_89}
\bibinfo{author}{\bibfnamefont{T.~J.} \bibnamefont{Thornton}},
  \bibinfo{author}{\bibfnamefont{M.~L.} \bibnamefont{Roukes}},
  \bibinfo{author}{\bibfnamefont{A.}~\bibnamefont{Scherer}}, \bibnamefont{and}
  \bibinfo{author}{\bibfnamefont{B.~P.} \bibnamefont{Van~de Gaag}},
  \bibinfo{journal}{Phys. Rev. Lett.} \textbf{\bibinfo{volume}{63}},
  \bibinfo{pages}{2128} (\bibinfo{year}{1989}).

\bibitem[{\citenamefont{Ford et~al.}(1989)\citenamefont{Ford, Washburn,
  B\"uttiker, Knoedler, and Hong}}]{Quenching_FORD_PRL89}
\bibinfo{author}{\bibfnamefont{C.~J.~B.} \bibnamefont{Ford}},
  \bibinfo{author}{\bibfnamefont{S.}~\bibnamefont{Washburn}},
  \bibinfo{author}{\bibfnamefont{M.}~\bibnamefont{B\"uttiker}},
  \bibinfo{author}{\bibfnamefont{C.~M.} \bibnamefont{Knoedler}},
  \bibnamefont{and} \bibinfo{author}{\bibfnamefont{J.~M.} \bibnamefont{Hong}},
  \bibinfo{journal}{Phys. Rev. Lett.} \textbf{\bibinfo{volume}{62}},
  \bibinfo{pages}{2724} (\bibinfo{year}{1989}).

\bibitem[{\citenamefont{Cheng et~al.}(2000)\citenamefont{Cheng, Yang, and
  Yang}}]{APL_InAs_RIE}
\bibinfo{author}{\bibfnamefont{K.~A.} \bibnamefont{Cheng}},
  \bibinfo{author}{\bibfnamefont{C.~H.} \bibnamefont{Yang}}, \bibnamefont{and}
  \bibinfo{author}{\bibfnamefont{M.~J.} \bibnamefont{Yang}},
  \bibinfo{journal}{Appl. Phys. Lett.} \textbf{\bibinfo{volume}{77}},
  \bibinfo{pages}{2861} (\bibinfo{year}{2000}).

\bibitem[{\citenamefont{Yang et~al.}(2002)\citenamefont{Yang, Yang, Cheng, and
  Culbertson}}]{InAS_PRB_2002}
\bibinfo{author}{\bibfnamefont{C.~H.} \bibnamefont{Yang}},
  \bibinfo{author}{\bibfnamefont{M.~J.} \bibnamefont{Yang}},
  \bibinfo{author}{\bibfnamefont{K.~A.} \bibnamefont{Cheng}}, \bibnamefont{and}
  \bibinfo{author}{\bibfnamefont{J.~C.} \bibnamefont{Culbertson}},
  \bibinfo{journal}{Phys. Rev. B} \textbf{\bibinfo{volume}{66}},
  \bibinfo{pages}{115306} (\bibinfo{year}{2002}).

\bibitem[{\citenamefont{Gilbertson et~al.}(2011)\citenamefont{Gilbertson,
  Fearn, Korm\'anyos, Read, Lambert, Emeny, Ashley, Solin, and
  Cohen}}]{InSb_PRB2011}
\bibinfo{author}{\bibfnamefont{A.~M.} \bibnamefont{Gilbertson}},
  \bibinfo{author}{\bibfnamefont{M.}~\bibnamefont{Fearn}},
  \bibinfo{author}{\bibfnamefont{A.}~\bibnamefont{Korm\'anyos}},
  \bibinfo{author}{\bibfnamefont{D.~E.} \bibnamefont{Read}},
  \bibinfo{author}{\bibfnamefont{C.~J.} \bibnamefont{Lambert}},
  \bibinfo{author}{\bibfnamefont{M.~T.} \bibnamefont{Emeny}},
  \bibinfo{author}{\bibfnamefont{T.}~\bibnamefont{Ashley}},
  \bibinfo{author}{\bibfnamefont{S.~A.} \bibnamefont{Solin}}, \bibnamefont{and}
  \bibinfo{author}{\bibfnamefont{L.~F.} \bibnamefont{Cohen}},
  \bibinfo{journal}{Phys. Rev. B} \textbf{\bibinfo{volume}{83}},
  \bibinfo{pages}{075304} (\bibinfo{year}{2011}).

\bibitem[{\citenamefont{Knez}(2012)}]{Knez_thesis}
\bibinfo{author}{\bibfnamefont{I.}~\bibnamefont{Knez}}, Ph.D. thesis,
  \bibinfo{school}{Rice University} (\bibinfo{year}{2012}).

\bibitem[{\citenamefont{Ihn}(2010)}]{Thomas_book}
\bibinfo{author}{\bibfnamefont{T.}~\bibnamefont{Ihn}},
  \emph{\bibinfo{title}{Semiconductor nanostructures}}
  (\bibinfo{publisher}{Oxford University Press New York},
  \bibinfo{year}{2010}).

\bibitem[{\citenamefont{Nichele}(2014)}]{Fabrizio_thesis}
\bibinfo{author}{\bibfnamefont{F.}~\bibnamefont{Nichele}}, Ph.D. thesis,
  \bibinfo{school}{ETH Zurich} (\bibinfo{year}{2014}).

\bibitem[{\citenamefont{Chaghi et~al.}(2009)\citenamefont{Chaghi, Cervera,
  Aït-Kaci, Grech, Rodriguez, and Christol}}]{Wet_etching_superlattice}
\bibinfo{author}{\bibfnamefont{R.}~\bibnamefont{Chaghi}},
  \bibinfo{author}{\bibfnamefont{C.}~\bibnamefont{Cervera}},
  \bibinfo{author}{\bibfnamefont{H.}~\bibnamefont{Aït-Kaci}},
  \bibinfo{author}{\bibfnamefont{P.}~\bibnamefont{Grech}},
  \bibinfo{author}{\bibfnamefont{J.~B.} \bibnamefont{Rodriguez}},
  \bibnamefont{and} \bibinfo{author}{\bibfnamefont{P.}~\bibnamefont{Christol}},
  \bibinfo{journal}{Semiconductor Science and Technology}
  \textbf{\bibinfo{volume}{24}}, \bibinfo{pages}{065010}
  (\bibinfo{year}{2009}).

\end{thebibliography}
\end{document}